\documentclass[aps,floatfix,onecolumn,a4paper,noshowpacs, nofootinbib,superscriptaddress,11pt]{revtex4}
\usepackage{graphicx,float}\usepackage{graphicx,float}
\usepackage[all]{xy}
\usepackage{amsmath,upgreek,bigints}
\usepackage{amssymb}
\usepackage{color}
\usepackage{epsfig,bm}		
\usepackage{graphicx,epstopdf}
\usepackage{subfigure}
\usepackage{pdfpages,slashed}
\usepackage[colorlinks,hyperindex]{hyperref}
\newcommand{\cvm}{current version of the manuscript}

\definecolor{green1}{RGB}{0,128,0}
\hypersetup{backref=true,pagebackref=true}
\hypersetup{%
  colorlinks = true,
  linkcolor  = blue,
  citecolor = blue
}
\usepackage{bookmark,textgreek}
\usepackage{hyperref,color,xcolor}
\hypersetup{hidelinks,hyperindex=true,colorlinks=true,breaklinks=true,urlcolor= blue}
\hypersetup{%
  colorlinks = true,
  linkcolor  = blue
}\usepackage{amssymb}
\newsavebox{\foobox}

\usepackage{graphicx,float,tikz}
\usepackage[all]{xy}
\newcommand\ringring[1]{%
  {
   \mathop{\kern0pt #1}\limits^{
     \vbox to-1.85ex{
       \kern-2ex 
       \hbox to 0pt{\hss\normalfont\kern.1em \r{}\kern-.45em \r{}\hss}%
       \vss 
     }
   }
  }
}
\newcommand\orcidgayane{{\href{https://orcid.org/0000-0002-0785-2826}{\orcidicon}}}
\newcommand{\orcidicon}{%
	\begin{tikzpicture}
	\draw[lime, fill=lime] (0,0)
		circle [radius=0.16]
		node[white] {{\fontfamily{qag}\selectfont \tiny ID}};
	\draw[white, fill=white] (-0.0625,0.095)
		circle [radius=0.007];
	\end{tikzpicture}	\hspace{-2mm}
}
\newcommand{\bpartial}{\mathop{\partial\kern -4pt\raisebox{.8pt}{$|$}}}

\newcommand{\bes}{\begin{subequations}}
\newcommand{\ees}{\end{subequations}}
\def\beq{\begin{eqnarray}}
 \newcommand{\clt}{\textcolor{black}}
  
\def\eeq{\end{eqnarray}}
\def\be{\begin{equation}}
\def\ee{\end{equation}}

\begin{document}

\title{Configurational entropy and the ${\cal N}^*(1440)$ Roper  resonance in QCD}
\author{G. Karapetyan\!\orcidgayane}
\email{gayane.karapetyan@ufabc.edu.br}
\affiliation{Federal University of ABC, Center of Mathematics, Santo Andr\'e, 09580-210, Brazil}
\affiliation{Perimeter Institute for Theoretical Physics,
Waterloo, Ontario, N2L 2Y5, Canada}

\begin{abstract}
The electroexcitation of the ${\cal N}^*(1440)$ Roper resonance, which defines the first radially excited state of the nucleon, is examined within the soft-wall AdS/QCD model. Such excited Fock states are characterized by the leading three-quark component, which determines the main properties of Roper resonance.
The differential configurational entropy (DCE) was used in the nuclear interaction with a gauge vector field for ${\cal N}^*(1440)$ transition. Comparing the main results with the recent data of the CLAS Collaboration at JLab shows a good agreement on the accuracy of the computed data.
\end{abstract}
\maketitle
\section{Introduction}
Nuclear resonances are an intriguing aspect of the anti-de Sitter (AdS)/quantum chromodynamics correspondence (QCD). Investigation of such phenomena, in particular, the Roper resonance, can shed new light on the mechanism ruling interactions among hadrons, the structure of the nucleons, as well as the fundamental features of the electromagnetic transitions between the nucleon and its resonances.
Within the soft-wall AdS/QCD model, one can determine the form factors and helicity amplitudes with high precision at a large range of $Q^2$, using the correct power scaling description.
The investigation of the nucleon resonances in the AdS/QCD soft wall can model the nucleon-Roper transition, where the Dirac form factor can be successfully determined by the holographic light-front QCD \cite{Brodsky:2006uqa,Erlich:2005qh,Karch:2006pv}. 
Recently, experiments of the CLAS Collaboration at JLab \cite{CLAS:2009ces} have shown the most prominent results on the electroexcitation of the low mass resonances. Such results were in a comprehensive analysis of data on differential cross sections, longitudinally polarized beam asymmetries, and longitudinal target and beam-target asymmetries for pion electroproduction off the proton.
The data obtained can be analyzed using different conceptual approaches.
Among both nonrelativistic and relativistic approaches to studying the electro-excitations of nucleon resonances, one can distinguish the potential and hadronic molecular approaches, Dyson-Schwinger equation framework, or the light-front holographic quantum QCD, \cite{Obukhovsky:2011sc,Aznauryan:2011qj,Aznauryan:2012ba,
Vega:2016gip,Gursoy:2007er,Ferreira:2019nkz,Brodsky:2003px,Mamo:2021krl,Hashimoto:2019wmg,Baldino:2021uie}.
As it has been marked in the above-mentioned studies, the Roper electroproduction requires additional degrees of freedom to be correctly interpreted for a nucleon-scalar $\sigma$ meson molecular component.
Also, in the framework of the light-front holographic QCD, the Dirac form factor for the electromagnetic nucleon-Roper transition was  calculated.
The dynamically generated resonance ${\cal N}^*(1440)$ suggests the estimation of the electromagnetic helicity form factors, besides valence quark contribution \cite{Ramalho:2017pyc}. 
The electroexcitation of resonances can be studied via the unitary isobar model MAID, which examines the photoproduction and electroproduction, as well as the nucleon resonances \cite{Drechsel:2007if}.

The Roper electroproduction, which comprises in addition of the leading three-quark ($3q$) state the higher Fock components either, has been investigated within a soft-wall AdS/QCD model \cite{Approach,Approach2,Gutsche:2012bp,daRocha:2022bnk}.
The AdS/QCD model for baryon structure \cite{Gutsche:2012bp} has been successfully used to investigate the nucleon electromagnetic form factors in the Euclidean region of transverse momentum squared, $p^2_T$, up to 30 GeV$^2$.
In such a study the higher Fock states, which have been incorporated beyond the $3q$ state allowed to describe properly the quantitative reproduction of baryons, i. e., their masses, radii, and the electromagnetic form factors at a small value of $Q^2$.
The Fock states have been limited by the twist dimension $\uptau=5$ and the higher states have been neglected. Such a choice is supported by the fact that at higher values of $Q^2$, the estimated contribution of higher Fock states to the hadronic form factors is as $(1/Q^2)^{\uptau-1}$.
In such cases, the number of adjusting parameters also decreases.
The mechanism of the ${\cal N}^*(1440)$ resonance with the determination of the helicity amplitudes at low-$Q^2$, takes into account the minimal coupling of the nucleon (with the orbital momentum $L=0$) and the ${\cal N}^*(1440)$ (with the orbital momentum $L=1$) Fock components with the same twist dimension, and leading twists for the nucleon and the ${\cal N}^*(1440)$ are $\uptau = 3$ and $\uptau = 4$, respectively.

To address the ${\cal N}^*(1440)$ transition, the differential configurational entropy  (DCE) will be used as a very important criterion to examine some of the most essential phenomenological features of AdS/QCD.
The DCE has been successfully applied to study phenomena in QCD   \cite{daRocha:2020gee,Fernandes-Silva:2018abr,Barreto:2022mbx,
daRocha:2021,daRocha:2021ntm,Karapetyan:2018yhm,daRocha:2021imz,daRocha:2021xwq,
Ferreira:2020iry,Bernardini:2019stn,daRocha:2021jzn,Bonora:2014dfa,Karapetyan:2022rpl}.
The DCE has been also used for studying different nuclear reactions, which involve high-energy hadrons under extremal circumstances \cite{Karapetyan:epjp,Karapetyan:plb,Karapetyan:2020epl,Karapetyan:2021epjp,Boschi-Filho:2005xct,dePaula:2008fp,Bernardini:2003ht}.
The latest studies in the such area have shown the useful prospect of DCE techniques to interpret QCD phenomenology \cite{Ferreira:2019inu,Bernardini:2018uuy,Braga:2018fyc,
Barbosa-Cendejas:2018mng,Barreto:2022len,Barreto:2022ohl}, also involving glueballs \cite{Bernardini:2016qit,MarinhoRodrigues:2020yzh,MarinhoRodrigues:2020ssq}, charmonium and bottomonium \cite{Braga:2017fsb}, the quark-gluon plasma \cite{daSilva:2017jay}, baryons \cite{Colangelo:2018mrt}, besides other aspects of information entropy in AdS/QCD \cite{Ma:2018wtw}.

Detailed information about the DCE can be found in Refs. \cite{Gleiser:2012tu, Gleiser:2011di, Gleiser:2013mga, Gleiser:2018kbq,Gleiser:2014ipa}.
Another set of studies is devoted to DCE phenomena  in AdS \cite{Casadio:2016aum,Fernandes-Silva:2019fez,Braga:2016wzx,Bernardini:2016hvx}.
Among the latest studies, which incorporate the DCE in the Color Glass Condensate (CGC) settings, one can mention the calculations of the inelastic hadron cross-section in deep hadron-hadron scattering
\cite{Karapetyan:2019epl,Karapetyan:2017edu}.
As the hadron cross-section represents the localized function of the probability of nuclear reaction in any spatial configuration of the excited system, we can calculate the critical points of the DCE using the Fourier transform of the cross-section and associated modal fraction
\cite{Karapetyan:2017edu,Karapetyan:2016fai}.

This paper is devoted to the investigation of the electromagnetic transitions between the nucleon and its resonances within the soft-wall AdS/QCD. We intend to study the Roper resonance in the negative-parity state ${\cal N}^*(1440)$.
We use the formalism suggested in Ref. \cite{Gutsche:2012wb}, which uses the adjustable quantum numbers for the gauge invariance for the form factors and helicity amplitudes momentum dependence. Such an approach allows determining the longitudinal $S_{1/2}$ amplitude in the low $Q^2$ regime. 
\clt{This manuscript emulates the recent results in Ref.  \cite{daRocha:2022bnk}, which implemented the DCE for the ${\cal N}^*(1535)$ resonance, to the ${\cal N}^*(1440)$ Roper resonance. The DCE approaches are similar, however each one of the resonances has peculiar physical features.  Here we originally provide the DCE approach, extending the results in Ref.  \cite{daRocha:2022bnk} to encompass the ${\cal N}^*(1440)$ Roper resonance.} This work is organized as follows:
Section II shortly represents the formalism of nucleon resonances in soft-wall AdS/QCD, with special attention to the Roper resonance  ${\cal N}^*(1440)$.
In Section III, we compute the DCE underlying the electroproduction of the Roper resonance and discuss the main results and analysis. 
In Section IV we present the summary and outlook of our research.

\section{Fermion field in AdS space}

In this paper, we use the AdS/QCD soft-wall holographic model, which describes the main properties of the electromagnetic structure of the nucleon, such as the analytical power scaling of the elastic nucleon form factors at large momentum transfers \cite{Gutsche:2012bp}. The model also reproduces satisfactorily the experimental data for magnetic moments and electromagnetic radii.
In AdS space, one can specify the five-dimensional AdS metric as
\begin{eqnarray}
\label{action1}
ds^2  = g_{MN} dx^M dx^N = 
\eta_{ab} \, \exp(2A(z)) \, dx^a dx^b
= \exp(2A(z)) \, (\eta_{\mu\nu} dx^\mu dx^\nu - dz^2)\,,\end{eqnarray}
with $
\eta_{\mu\nu} = {\rm diag}(1, -1, -1, -1, -1).$ 
In Eq. (\ref{action1}) the indexes $a=(\mu,z)$ and $b=(\nu,z)$ specify local Lorentz indexes, and $g_{MN}$ and  $\eta_{ab}$ are curved and flat metric tensors, respectively, which are related by the vielbein $\varepsilon_M^a(z)= exp(A(z)) \, \delta_M^a$ as  $g_{MN} =\varepsilon_M^a \varepsilon_N^b \eta_{ab}$, with
$g = |{\rm det} g_{MN}| = \exp(10 A(z))$, $R$ is the AdS radius, $z$ is the holographic coordinate, and
$M,N = 0, 1, \ldots, 4$ are the base manifold space-time indices.
In AdS space the calculation approach is restricted by a conformal-invariant metric with warp factor $A(z) = \log(R/z)$,
and the corresponding AdS/QCD action for the fermion field of twist $\uptau$ is given by an expression \cite{Gutsche:2012bp}:
\begin{eqnarray}
\label{eff_action}
S_\uptau = \int d^4x dz \, \sqrt{g} \, \exp(-\upphi(z)) \, \sum\limits_{i=+,-}
\bar\uppsi_{i,\uptau}(x,z)
\, \mathfrak{D}_i(z) \uppsi_{i,\uptau}(x,z) \,,
\end{eqnarray}
with covariant derivative
\begin{eqnarray}
\mathfrak{D}_\pm(z) =  \frac{i}{2} \upgamma^M
\! \stackrel{\leftrightarrow}{\partial}_{_M}
\, \mp \,  (\mu + U_F(z))\,,\label{ac1}
\end{eqnarray}
where the parameter $\mu$ represents the bulk fermion mass associated to the scaling dimension $\uptau$ ($m = \mu R = \uptau - 3/2$),
$\uppsi_{\pm,\uptau}(x,z)$ is the pair of bulk fermion fields, and in the 4D theory, they are related to the holographic analogs of the left- and right-chirality operators. The dilaton field is assumed in its standard quadratic form,  $\upphi(z) = \kappa^2 z^2$, with $\kappa$ being a free scale parameter, and $\upgamma^M = \varepsilon_a^M \upgamma^a$ and  $\upgamma^a=(\upgamma^\mu, - i\upgamma^5)$ are the five-dimensional Dirac matrices in the chiral representation for the $\upgamma^\mu$ and $\upgamma^5$ matrices. 
The fields $\uppsi_\uptau$ determine the AdS fermion field for the different scaling dimension: $\uptau = 3, 4, 5, \ldots$.
In the holographic QCD, the scaling dimension of the baryon interpolating operator in the light-front wave, $\uptau = N + L$, can be suggested through the scaling dimension of the AdS fermion field, where $L = {\rm max} \, | L_z |$ is the maximal value of the $z$- component of the quark orbital angular momentum, and $N$ is the number of partons in the baryon \cite{Brodsky:2006uqa}.
The effective potential $U_F(z) = \upphi(z)/R$ must satisfy the solutions of the motion equations (EOMs) for the fermionic Kaluza-Klein (KK) modes of left- and right-chirality, and correctly describe the asymptotic behavior of the nucleon electromagnetic form factors at large $Q^2$.
The $P$-parity transformation condition implies the fermion masses $m$ and the effective potentials $U_F(z)$ for $\uppsi_+$ and $\uppsi_-$  fields must have the opposite signs \cite{Gutsche:2012bp}, and the absolute sign of the fermion mass depends on the chirality of the boundary operator \cite{Hong:2006ta}.
Within the present approach one can choose for the QCD operators ${\cal O}_R$ and ${\cal O}_L$ positive and negative chirality. Thus, the correspondent absolute $\pm$ signs have the mass terms of the bulk fields $\uppsi_\pm$.

Within the soft-wall AdS/QCD model, one can calculate some of the parameters that regulate wave functions of the interacting hadrons. The first step is to rescale the fermionic fields, using the dilaton, as the following \cite{Gutsche:2012wb}
\begin{eqnarray}
\uppsi_{i, \uptau}(x,z) \mapsto \exp(\upphi(z)/2) \uppsi_{i, \uptau}(x,z).
\end{eqnarray}
as well as to remove the dilaton field from the overall exponential. 
It is possible to represent the modified action in the Lorentzian signature via the $\uppsi_\uptau(x,z)$ field as:
\begin{eqnarray}
\label{S_F2}
\begin{split}
S_\uptau \!=\! \int d^4x dz  \exp(4A(z))  \sum\limits_{i=\pm}
\bar\uppsi_{i,\uptau}(x,z)\!
\biggl\{ i\slashed\partial \!+\! \upgamma^5\partial_z 
\!+\! 2 A^\prime(z) \upgamma^5
\!-\!  \delta_i \frac{\exp(A(z))}{R} \Big(m \!+\! \upphi(z)\Big)  \biggr\}
 \uppsi_{i, \uptau}(x,z)\,
\end{split}
\end{eqnarray}
with $\not\!\partial = \upgamma^\mu \, \partial_\mu$, $\delta_\pm = \pm 1$, and the fermion field, $\uppsi_{i, \uptau}(x,z)$, must satisfy the equations of motion in the form \cite{Gutsche:2012bp}
\begin{eqnarray}
\biggl[ i\not\!\partial + \upgamma^5\partial_z
+ 2 A^\prime(z) \upgamma^5
\mp \frac{\exp(A(z))}{R} \Big(m + \upphi(z)\Big) \biggr]
\uppsi_{\pm,\uptau}(x,z) = 0\,.
\end{eqnarray}
In the soft-wall AdS/QCD model, the fermion field can be split into the left- (L) and right-chirality (R) components \cite{Gutsche:2012wb}:
\begin{eqnarray}
\uppsi_{i, \uptau}(x,z) &=& \uppsi^L_{i,\uptau}(x,z) + \uppsi^R_{i, \uptau}(x,z)\,, \quad
\\
\uppsi^{L/R}_{i, \uptau}(x,z) &=&
\frac{1 \mp \upgamma^5}{2} \uppsi_{i, \uptau}(x,z) \,,
\end{eqnarray}
where $\mathsf{F}^{L/R}_{i, \uptau,n}(z)$ are the profile
functions, and
$\uppsi^{L/R}_{i, \uptau}(x,z)$ fields
can be represented through the four-dimensional boundary fields in the fermionic KK modes as:
\begin{eqnarray}
\uppsi^{L/R}_{i, \uptau}(x,z) = \frac{1}{\sqrt{2}} \, \sum\limits_n
\ \uppsi^{L/R}_n(x) \ \mathsf{F}^{L/R}_{i, \uptau, n}(z) \,.
\end{eqnarray}
The left- and right-chirality components also compose the Dirac (bi)spinors $\uppsi_n(x) = \uppsi^L_n(x) + \uppsi^R_n(x)$.
The profile functions are linked owing to the four-dimensional $P$- and $C$-parity invariance as the following \cite{Gutsche:2012bp},
\begin{eqnarray}
\mathsf{F}^R_{\pm, \uptau, n}(z) = \pm \mathsf{F}^L_{\mp, \uptau, n}(z)\,.
\end{eqnarray}
It is more convenient to use the alternative definition
\begin{eqnarray}
\begin{split}
\mathsf{F}^R_{\uptau, n}(z) \equiv
\mathsf{F}^R_{+, \uptau, n}(z) = \mathsf{F}^L_{-, \uptau, n}(z)\,,\\
\mathsf{F}^L_{\uptau, n}(z) \equiv
\mathsf{F}^L_{+, \uptau, n}(z) = - \mathsf{F}^R_{-, \uptau, n}(z)\,.
\end{split}
\end{eqnarray}
The AdS/QCD model profile functions $\mathsf{F}^{L/R}_{\uptau, n}(z)$ are the holographic analogs of the nucleon wave function and are defined by the quantum number $n$ and twist dimension $\uptau$, which can be considered as the specific partonic content of the nucleon Fock component.
The profile functions satisfy the two coupled one-dimensional motion equations \cite{Gutsche:2012bp} in the form:
\begin{eqnarray}\label{sa1}
\biggl[\partial_z \pm \frac{\exp(A)}{R} \, \Big(m+\upphi(z)\Big)
+ 2 A^\prime \biggr] \mathsf{F}^{L/R}_{n, \uptau}(z) = \pm M_{n\uptau}
\mathsf{F}^{R/L}_{n, \uptau}(z) \,.
\end{eqnarray}

In order to find the solutions for the AdS field profiles in the $z$ direction, one obtains the decoupled EOMs as:
\begin{eqnarray}
\begin{split}
\biggl[\partial_z^2 \!+\! 4 A^\prime \partial_z
\!-\! \frac{\exp(2A)}{R^2} (m+\upphi)^2
\mp \frac{\exp(A)}{R}
\Big(A^\prime (m\!+\!\upphi) \!-\! \upphi^\prime\Big)
\!+\! 4 A^{\prime 2} \!+\! 2 A^{\prime\prime}+M_{n\uptau}^2
\biggr] \mathsf{F}^{L/R}_{\uptau, n}(z) \!=\!0 \,.
\end{split}
\end{eqnarray}
If we substitute $\mathsf{F}^{L/R}_{\uptau, n}(z) = \exp(- 2 A(z)) \, \mathsf{f}^{L/R}_{\uptau, n}(z)$,
then the Schr\"odinger-type equations of motion  for $\mathsf{f}^{L/R}_{\uptau, n}(z)$ can be expressed as:
\begin{eqnarray}
\label{eq_KK}
\begin{split}
\biggl[ -\partial_z^2
+ \frac{\exp(2A)}{R^2} (m+\upphi)^2
\mp \frac{\exp(A)}{R} \Big(A^\prime (m+\upphi) +
 \upphi^\prime\Big) \biggr] 
\mathsf{f}^{L/R}_{\uptau, n}(z) =
M_{n\uptau}^2 \, \mathsf{f}^{L/R}_{\uptau, n}(z) \,.
\end{split}
\end{eqnarray}
Then, substituting $A(z)=\log(R/z)$ and $\upphi(z)=\kappa^2 z^2$, then Eq.(\ref{eq_KK}) yields
\begin{eqnarray}
\begin{split}
\biggl[ -\partial_z^2
+ \kappa^4 z^2 + 2 \kappa^2 \Big(m \mp \frac{1}{2} \Big)
+ \frac{m (m \pm 1)}{z^2} \biggr] \mathsf{f}^{L/R}_{\uptau, n}(z) = 
M_{n\uptau}^2 \, \mathsf{f}^{L/R}_{\uptau, n}(z),
\end{split}
\end{eqnarray}
with the corresponding notations:
\begin{eqnarray}
\mathsf{f}^L_{\uptau, n}(z) &=& \sqrt{\frac{2\upgamma(n+1)}{\upgamma(n+\uptau)}}
\ \kappa^{\uptau}
\ z^{\uptau-1/2} \exp(-\kappa^2 z^2/2)  L_n^{\uptau - 1}(\kappa^2z^2) \,,\label{sa2} \\
\mathsf{f}^R_{\uptau, n}(z) &=& \sqrt{\frac{2\upgamma(n+1)}{\upgamma(n+\uptau-1)}}
\ \kappa^{\uptau-1} z^{\uptau-3/2} \exp({-\kappa^2 z^2/2}) L_n^{\uptau-2}(\kappa^2z^2),\label{sa3}
\end{eqnarray}
where $L_n^\uptau(x)$ are the generalized Laguerre polynomials and the mass spectrum $M_{n\uptau}$ satisfies Regge-like trajectories given by \begin{eqnarray}
M_{n\uptau}^2 = 4 \kappa^2 \Big( n + \uptau - 1 \Big),
\end{eqnarray}
for $m = \uptau  -  3/2$.
When we consider the profile function as the nucleon wave function of the nucleon $\mathsf{F}^{L/R}_{\uptau, n}(z) = \exp({- 2 A(z)})  \mathsf{f}^{L/R}_{\uptau, n}(z)$, then for small values of $z$, a correct scaling relation can be observed for both profile functions for the twist $\uptau$:
\begin{eqnarray}
\mathsf{F}^L_{\uptau, n}(z) \sim z^{\uptau+3/2}\,, \quad\quad
\mathsf{F}^R_{\uptau, n}(z) \sim z^{\uptau+1/2}\,,
\end{eqnarray}
and such a process vanishes during confinement at large values of $z$.
Within the soft-wall AdS/QCD model, bulk fields can play the role of  the holographic analog
of the nucleon resonance, $\uppsi^{\cal N}_{\pm,\uptau}(x,z)$, for the nucleon in the ground state with the quantum number $n=0$,
and holographic analog of the Roper resonance, $\uppsi^{\cal R}_{\pm,\uptau}(x,z)$, in the first radially excited state with $n=1$ \cite{Gutsche:2012wb}. They can be determined in the following form:
\begin{eqnarray}
\uppsi^{\cal N}_{\pm,\uptau}(x,z) &=& \frac{1}{\sqrt{2}} \,
\left[
      \uppsi^{L}_0(x) \ \mathsf{F}^{L/R}_{\uptau, 0}(z)
\pm   \uppsi^{R}_0(x) \ \mathsf{F}^{R/L}_{\uptau, 0}(z)\right]\,,\\
\uppsi^{\cal R}_{\pm,\uptau}(x,z) &=& \frac{1}{\sqrt{2}} \,
\left[
      \uppsi^{L}_1(x) \ \mathsf{F}^{L/R}_{\uptau, 1}(z)
\pm   \uppsi^{R}_1(x) \ \mathsf{F}^{R/L}_{\uptau, 1}(z)\right]\,.
\end{eqnarray}
We consider  AdS bulk fields as the products of boundary fermionic  fields with spin $1/2$, whose profiles depend on the holographic variable.
Once  the nucleon and Roper resonances are defined, one can represent the free actions at the fixed twist dimension, $\uptau$, as:
\begin{eqnarray}
S_\uptau^B = \int d^4x dz \exp(4A(z))  \mathcal{L}^B, 
\end{eqnarray}
where 
\beq\label{l123}
\!\!\!\mathcal{L}_\uptau^B=\sum\limits_{j=\pm}
\, \bar\uppsi^B_{j,\uptau}(x,z) \,
\biggl\{ i\not\!\partial + \upgamma^5\partial_z + 2 A^\prime(z) \upgamma^5
-  \delta_j \frac{\exp(A(z))}{R} \Big(m + \upphi(z)\Big)  \biggr\}
\, \uppsi^B_{j, \uptau}(x,z)\,,
\eeq
with the notation for nucleon and Roper resonances $B = {\cal N}, {\cal R}$, respectively.
Summing the bulk action, $S_\uptau^B$, over $\uptau$ values with adjustable coefficients $c_\uptau^B$ allows us to consider the higher Fock states for the nucleon and Roper resonances in the form:
\begin{eqnarray}
S^B = \sum\limits_\uptau \, c_\uptau^B \, S_\uptau^B \,,
\end{eqnarray}
where $c_\uptau^B$ $(B = {\cal N}, {\cal R})$ are adjustable parameters.
It is equivalent to expressing the effective Lagrangian
\begin{eqnarray}
\mathcal{L}^B = \sum\limits_\uptau \, c_\uptau^B \, \mathcal{L}_\uptau^B \,,\label{lag13}
\end{eqnarray}
weighted by the coefficients $c_\uptau^B$ regulating the nucleon and Roper resonances. Eq. (\ref{lag13}) will play a prominent role on
determining the DCE underlying the Roper resonance $\mathcal{N}^*(1440)$ in the next section, pointing to the values of $c_\uptau^B$ which correspond to higher configurational stability.

According to the electromagnetic gauge invariance, one uses the normalization of the kinetic term $\bar\uppsi_n(x) i\!\!\not\!\partial\uppsi_n(x)$ of the boundary fermionic field as $
\sum_\uptau \, c_\uptau^B = 1.$ 
The masses for the nucleon and Roper can be expressed as \cite{Gutsche:2012bp}:
\begin{eqnarray}
M_{\cal N} &=& 2 \kappa \sum\limits_\uptau \, c_\uptau^{\cal N} \, \sqrt{\uptau - 1}
\,, \qquad\qquad
M_{\cal R} = 2 \kappa \sum\limits_\uptau \, c_\uptau^{\cal R} \, \sqrt{\uptau} \;.
\end{eqnarray}
The four-dimensional actions for the fermion field $\uppsi_n(x) = \uppsi^L_n(x) + \uppsi^R_n(x)$ in
$n=0$ nucleon state and $n=1$ state for Roper resonance can be obtained by the integration procedure over the holographic coordinate $z$, yielding  
\begin{eqnarray}
S^B_{4D} \ = \ \int d^4x \, \mathcal{L}^B_{4D}, \end{eqnarray}
where
\beq
 \mathcal{L}^B_{4D} = \bar\uppsi_0(x) \biggl[ i \not\!\partial - M_{\cal N} \biggr] \uppsi_0(x) +
\bar\uppsi_1(x) \biggl[ i \not\!\partial - M_{\cal R} \biggr] \uppsi_1(x)
\,.\eeq
One can see that the expression for the effective actions in 4D is obtained using the bulk fields propagating in AdS. The additional dimension of the conventional hadrons is related to the baryon mass $M_B$.
For both, the nucleon and the Roper resonances, the model suggests the contribution of Fock states with twist $\uptau=3, 4$, and $5$.
Using the following values for
the parameters $\kappa$, $c_3^{\cal N}$, and $c_4^{\cal N}$ \cite{Gutsche:2012wb}
\begin{eqnarray}
\label{par1}
\kappa = 383 \ \mathrm{MeV}\,, \quad c_3^{\cal N} = 1.25\,, \quad c_4^{\cal N} = 0.16\,,
\end{eqnarray}
one can obtain the data for the nucleon/proton mass
$M_{\cal N}^{\rm exp} = 938.27$ MeV.
The parameter $c_5^{\cal N}$ related parameters $c_3^{\cal N}$ and $c_4^{\cal N}$ in the form:
\begin{eqnarray}
c_5^{\cal N} = 1 - c_3^{\cal N} - c_4^{\cal N} = - 0.41 \,.
\end{eqnarray}
In order to reproduce the Roper mass $M_{\cal R}^{\rm exp} = 1440$ MeV, we use the set of parameters \cite{Gutsche:2012wb}:
\begin{eqnarray}
\label{par2}
c_3^{\cal R} = 0.78\,, \quad  c_4^{\cal R} = - 0.16\,, \quad
c_5^{\cal R} = 1 - c_3^{\cal R} - c_4^{\cal R} = 0.38 \,.
\end{eqnarray}
One can conclude that the value of mass for both resonances is determined by the contribution of the $3q$ Fock component.

\section{Differential configurational entropy and the electro-excitation of the Roper resonance}

The main procedure to obtain the DCE assumes a probability distribution in the form of the localized energy density operator, which represents the temporal $T_{00}({\vec{r}})$ component of the energy-momentum tensor, for ${\vec{r}}=(x_1,\ldots,x_k)\in\mathbb{R}^k$.
 The Fourier transform of the energy density operator can be obtained as \cite{Gleiser:2018kbq}
\begin{eqnarray}
\label{fou}
{\tau_{00}({\vec{q}})} = \frac{1}{(2\pi)^{k/2}}\int_{\mathbb{R}^k}\,T_{00}({\vec{r}})e^{-i{\vec{q}}\cdot {\vec{r}}}\,{\rm d}^k x,
\end{eqnarray}
where {$\tau_{00}({\vec{q}})$} is the spatial part of the 4-momentum $q$. 
 One can define the modal fraction as \cite{Gleiser:2012tu}
\begin{eqnarray}
{\xi_{00}({\vec{q}})} =
\frac{\left|{\tau_{00}({\vec{q}})}\right|^{2}}{ \bigintsss_{\mathbb{R}^k} \,\left|{\tau_{00}(\mathsf{q})}\right|^{2}\,{\rm d}^k\mathsf{q}}.
\label{modalf}
\end{eqnarray}
The modal fraction (\ref{modalf}) reflects the relative weight transferred by each momentum wave mode ${\vec{q}}$.
The  DCE allows us to compute the energy density, which involves the complete information about the nuclear system described by the weight of each wave mode  \cite{Gleiser:2018kbq}:
\begin{eqnarray}
{\rm DCE}_{T_{00}}= - \int_{\mathbb{R}^k}\,{{\xi_{00}^{\scalebox{.57}{$\,\otimes$}}}({\mathsf{q}})}\ln \left({\xi_{00}^{\scalebox{.57}{$\,\otimes$}}}({\mathsf{q}})\right)\,{\rm d}^k\mathsf{q},
\label{confige}
\end{eqnarray}
with
\begin{eqnarray}
{\xi_{{00}}^{\scalebox{.57}{$\,\otimes$}}({\vec{q}})=\frac{\xi_{00}({\vec{q}})}{\xi_{{00}}^{\scalebox{.58}{max}}({\vec{q}})}},
\end{eqnarray}
and ${\xi_{{00}}^{\scalebox{.58}{max}}({\vec{q}})} = 3.451$ designating the maximum value of the energy density in the momentum space $\mathbb{R}^k$. 
Here we use the nat (natural unit of information) as the DCE unit since it defines the amount of information encoding in the probability distribution function with the equally likely outcomes in the interval $[0,e]$.

{For $k=1$ in (\ref{fou}) -- (\ref{confige}), corresponding to the integration along the holographic $z$ coordinate, one can calculate the DCE for the closed nuclear system over all values of $z$} and due to the Kaluza--Klein splitting. Using the 
Lagrangian density (\ref{lag13}), the energy density operator can be obtained as the temporal component of the energy-momentum tensor:
\begin{eqnarray}
{ \!\!\!\!\!\!\!\!T_{00}\!=\! \frac{2}{\sqrt{ -g }}\!\! \left[\frac{\partial (\sqrt{-g}{\mathcal{L}^B})}{\partial{g_{00}}} \!-\!\frac{\partial}{\partial{ x^\ell }} \frac{\partial (\sqrt{-g} {\mathcal{L}^B})}{\partial\left(\frac{{\scalebox{.79}{$\,\partial$}} g_{00}}{{\scalebox{.79}{$\,\partial$}}x^\ell}\right)}{+\partial_0\bar\psi\frac{\partial {\mathcal{L}^B}}{\partial\left(\partial^0\bar\psi\right)}+\frac{\partial {\mathcal{L}^B}}{\partial\left(\partial^0\psi\right)}\partial_0\psi - g_{00}{\mathcal{L}^B}},
 \right]},
\label{em1}
\end{eqnarray}
where $\psi$ accounts for any fermionic field entering in the Lagrangian \eqref{l123}.
Explicitly, Eq. (\ref{em1}) can be written as \beq
\label{actionS55}
T_{00}\! &\!=\!&\!\!\sum\limits_{\uptau; \;j=\pm} \clt{c_\uptau^B}\left[
\bar\uppsi^B_{j,\uptau}(x^\mu,z) \, \gamma_0{}\biggl\{ i\overset{\leftrightarrow}{\not\!\partial} + \upgamma^5\overset{\leftrightarrow}{\partial_z} + 2 A^\prime(z) \upgamma^5
-  \delta_j \frac{\exp(A(z))}{R} \Big(m + \upphi(z)\Big)  \biggr\} \,  \uppsi^B_{j, \uptau}(x,z)\right.\nonumber\\
&&\left.-g_{00} \bar\uppsi^B_{j,\uptau}(x,z) \,
\biggl\{ i{\not\!\partial} + \upgamma^5\partial_z + 2 A^\prime(z) \upgamma^5
-  \delta_j \frac{\exp(A(z))}{R} \Big(m + \upphi(z)\Big)  \biggr\}
\, \uppsi^B_{j, \uptau}(x^\mu, z)\right].\eeq

This procedure allows us to obtain the Fourier transform of the energy density operator, the modal fraction, and the DCE, respectively using Eqs. (\ref{fou}) -- (\ref{confige}).
The Lagrangian densities can be thought of as being functions of the parameters $c_3^{\cal N}$, $c_4^{\cal N}$ and $c_3^{\cal R}$, $c_4^{\cal R}$ in (\ref{par1}, \ref{par2}). Their phenomenological values in (\ref{par1}, \ref{par2}) were derived via the fitting procedure to the helicity amplitudes of the ${\cal N}^*(1440)$ resonance \cite{Gutsche:2012bp}.
Instead of using the values of $c_3^{\cal N}$, $c_4^{\cal N}$ and $c_3^{\cal R}$, $c_4^{\cal R}$ in (\ref{par1}, \ref{par2}), these values will be derived out of the global minima of the DCE, which indicates the maximum configurational stability of the nuclear system and the dominant state occupied by the nuclear physical system.
The minimal value of the DCE means that the nuclear system has a definite, well-localized energy density with small uncertainty.

\begin{figure}[H]
 \centering
 \includegraphics[width=4.5in]{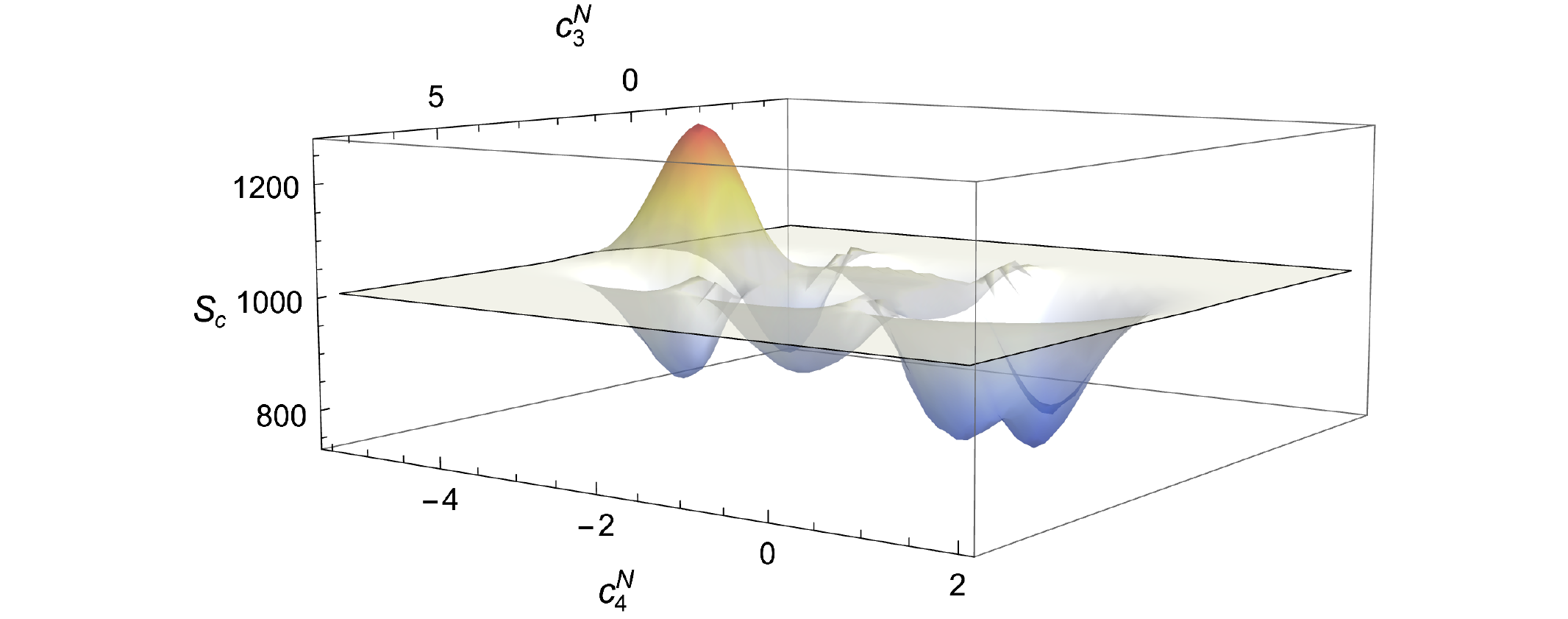}
 \caption{DCE ($S_c$) as a function of $c_3^{\cal N}$ and $c_4^{\cal N}$. There is a global minimum DCE$_{\textsc{min}}(c_3^{\cal N}, c_4^{\cal N}) = 767.503$ nat, for $c_3^{\cal N}= 1.264$ and $c_4^{\cal N} = 0.169$, respectively within an accuracy of 1.1\% and 5.6\%, comparing to data from the AdS/QCD soft-wall model in Refs. \cite{Gutsche:2012wb,Gutsche:2012bp}.}
 \label{plot1}
\end{figure}

\begin{figure}[H]
 \centering
 \includegraphics[width=3.1in]{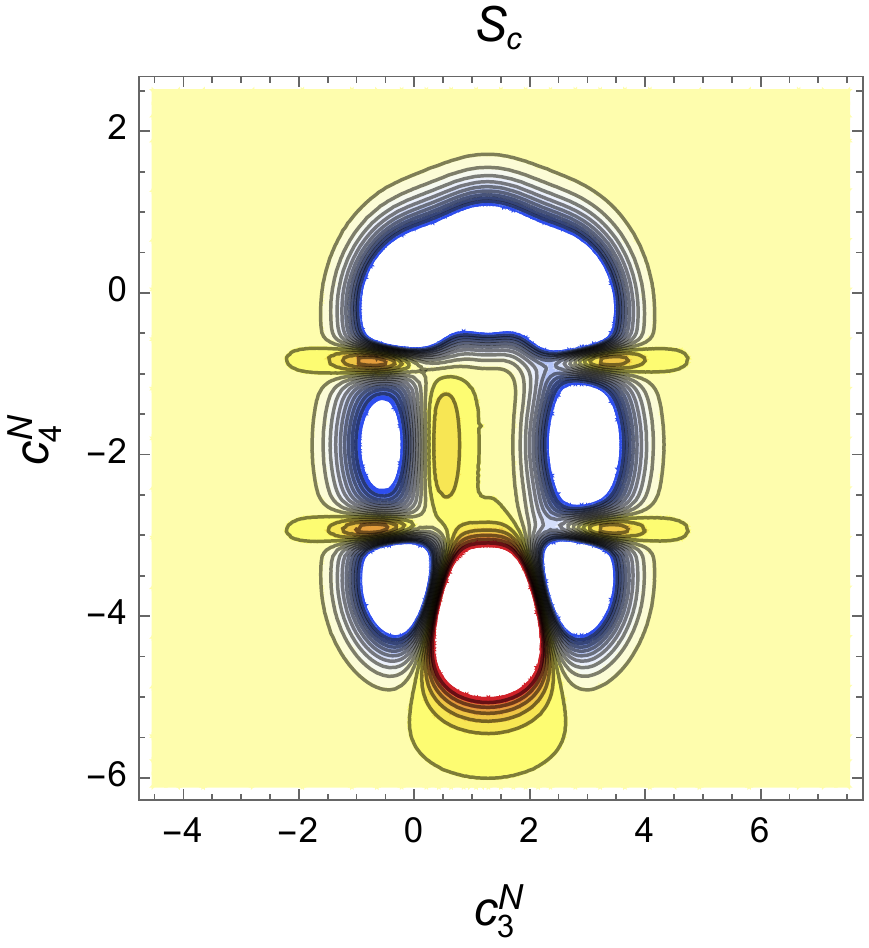}
 \caption{Contour plot of the
 DCE as a function of the parameters $c_3^{\scalebox{.6}{\textsc{{$\cal N$}}}}$ and $c_4^{\scalebox{.6}{\textsc{{$\cal N$}}}}$ for the nucleon resonance.}
 \label{plot2}
\end{figure}

In order to calculate the main properties of the ${\cal N}^*(1440)$ for the nucleon/Roper resonances, we fix two parameters, $\kappa$ = 383 MeV and the Roper resonance mass, $M_{\cal R}^{\rm exp} = 1440$ MeV. Therefore, we can determine two free adjustable parameters, $c_3^{\cal N}$, and $c_4^{\cal N}$; $c_3^{\cal R}$ and  $c_4^{\cal R}$, respectively in Eqs. (\ref{par1}, \ref{par2}), considering the other two complementary parameters fixed. Such a procedure allows us to obtain the global minimum of the DCE ($S_c$), hence, determining the most dominant state occupied by the nuclear quantum system.
We represent the results of the calculations of the DCE using Eqs. (\ref{fou}) -- (\ref{confige}) in Figs. \ref{plot1} and \ref{plot2} as a function of 
 the free parameters $c_3^{\cal N}$ and $c_4^{\cal N}$.

In Fig. \ref{plot2} the contour plots show the DCE as a function of parameters $c_3^{\scalebox{.6}{\textsc{{$\cal N$}}}}$, $c_4^{\scalebox{.6}{\textsc{{$\cal N$}}}}$  for the nucleon, respectively.
The contour plot in Fig. \ref{plot2} illustrates the DCE by slices of 
isentropic domains. 
As one can see from Fig. \ref{plot2} for the nucleon resonance, the dark center of the yellow domain of closed curves indicates the point (\ref{dce11}), which matches the global minimum of the DCE (\ref{dce1}).
\noindent The obtained results for the nucleon resonance show that the global minimum (\textsc{gmin})
\begin{eqnarray}
\text{DCE$\left(c_{3\,{\scalebox{0.6}{$\textsc{gmin}$}}}^{\scalebox{.6}{{$\cal N$}}}, c_{4\,{\scalebox{0.6}{$\textsc{gmin}$}}}^{\scalebox{.6}{{$\cal N$}}}\right)$= 767.503 nat},
\label{dce1}
\end{eqnarray}
at
\begin{eqnarray}
 \text{$c_{3\,{\scalebox{0.6}{$\textsc{gmin}$}}}^{\scalebox{.6}{\textsc{{$\cal N$}}}}=1.264$, \qquad\qquad $c_{4\,{\scalebox{0.6}{$\textsc{gmin}$}}}^{\scalebox{.6}{\textsc{{$\cal N$}}}} =0.169$},
\label{dce11}
\end{eqnarray} match the data in Eq. (\ref{par1}) from the AdS/QCD soft-wall model in Refs. \cite{Gutsche:2012wb,Gutsche:2012bp} with an accuracy of 1.1\% and 5.6\%, respectively.
In order to run over all possible values of adjustable parameters, corresponding to the experimental nucleon masses and the Roper mass, the range of calculation for $c_3^{\scalebox{.6}{\textsc{{$\cal N$}}}}$,  $c_4^{\scalebox{.6}{\textsc{{$\cal N$}}}}$  has been chosen wide enough.

\begin{figure}[H]
 \centering
 \includegraphics[width=4.5in]{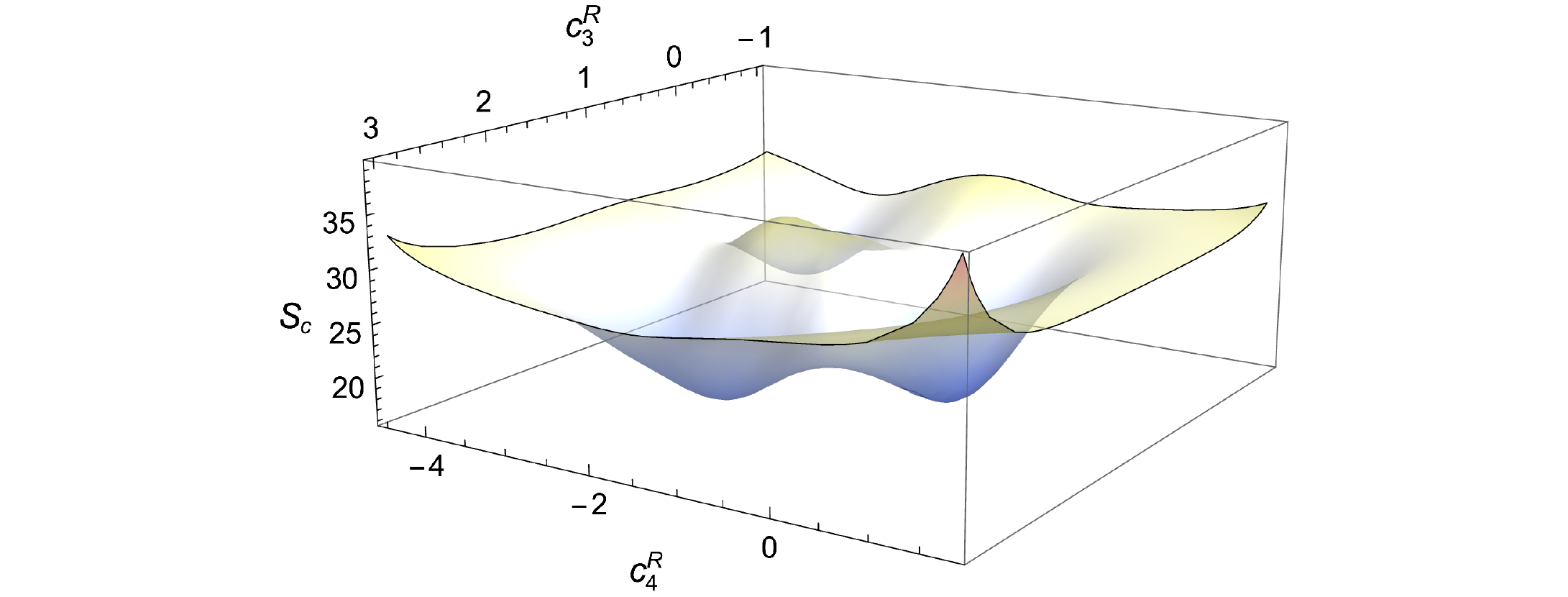}
 \caption{DCE ($S_c$) as a function of $c_3^{\cal R}$ and $c_4^{\cal R}$. There is a global minimum DCE$_{\textsc{min}}(c_3^{\cal R}, c_4^{\cal R}) = 17.639$ nat, for $c_3^{\cal R}= 0.803$ and $c_4^{\cal R} = -0.171$, respectively within an accuracy of 2.9\% and 6.8\%, comparing to data from the AdS/QCD soft-wall model in Refs. \cite{Gutsche:2012wb,Gutsche:2012bp}.}
 \label{plot3}
\end{figure}

\begin{figure}[H]
 \centering
 \includegraphics[width=3.1in]{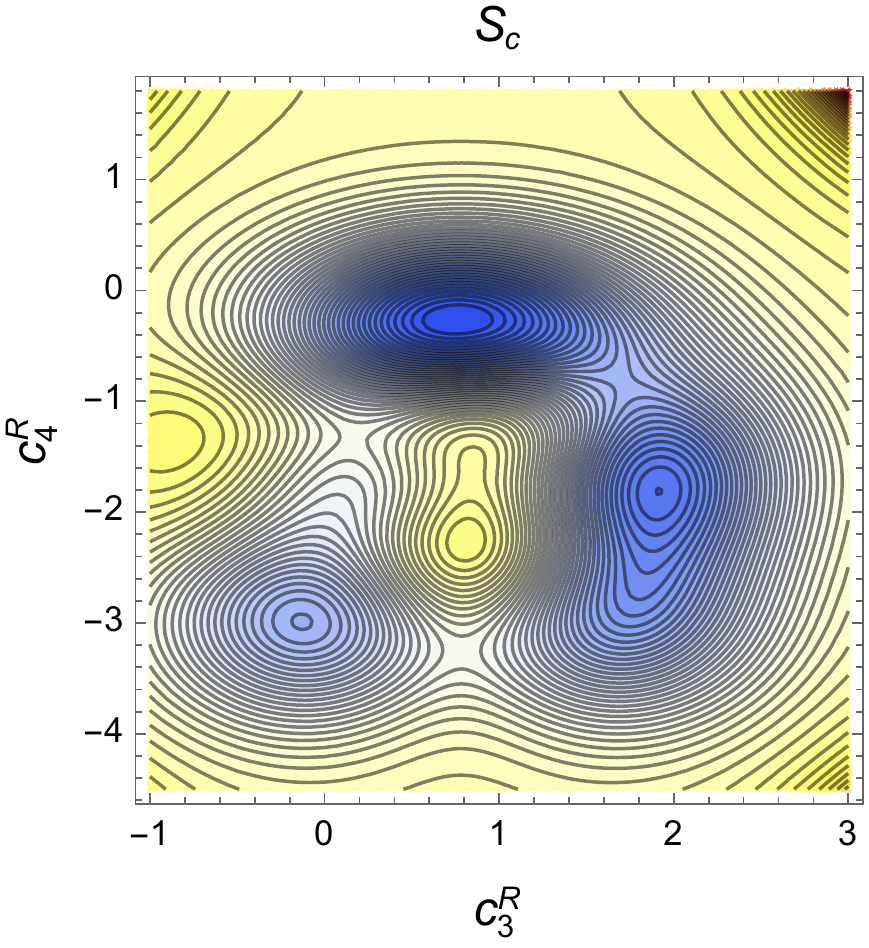}
 \caption{Contour plot of the
 DCE as a function of the parameters $c_3^{\scalebox{.6}{\textsc{$\cal R$}}}$ and $c_4^{\scalebox{.6}{\textsc{$\cal R$}}}$ for the Roper resonance.}
 \label{plot4}
\end{figure}

Now, the calculations of the DCE using Eqs. (\ref{fou}) -- (\ref{confige}) are depicted in Figs. \ref{plot3} and \ref{plot4} as a function of 
 the free parameters  $c_3^{\cal R}$ and $c_4^{\cal R}$, where the prevailing quantum states are reached through the global minimum of the DCE.
For the Roper resonance, Fig. \ref{plot3} reveals the DCE global minimum 
\begin{eqnarray}
\text{DCE$\left(c_{3\,{\scalebox{0.6}{$\textsc{gmin}$}}}^{\scalebox{.6}{\textsc{$\cal R$}}}, c_{4\,{\scalebox{0.6}{$\textsc{gmin}$}}}^{\scalebox{.6}{\textsc{$\cal R$}}}\right)$ = 17.639 nat},
\label{dce10}
\end{eqnarray}
at the point of the parameter space
\begin{eqnarray}
 \text{$c_{3\,{\scalebox{0.6}{$\textsc{gmin}$}}}^{\scalebox{.6}{\textsc{$\cal R$}}}=0.803$, \qquad\qquad $c_{4\,{\scalebox{0.6}{$\textsc{gmin}$}}}^{\scalebox{.6}{\textsc{$\cal R$}}} =-0.171$},
\label{dce110}
\end{eqnarray}
matching the data from the AdS/QCD soft-wall model in Ref. \cite{Gutsche:2012wb,Gutsche:2012bp}, given by Eq. (\ref{par2}), with an accuracy of 2.9\% and 6.8\%, respectively.

The fitted parameters $(c_3^{\scalebox{.6}{\textsc{$\cal R$}}}$, $c_4^{\scalebox{.6}{\textsc{$\cal R$}}}$) completely define the parameter space, in which the Roper resonance acquires the configurational stability of a higher level at the global minimum, which is also the most dominant state among all possible quantum states.
Our data, corresponding to the global minimum for the nucleon resonance DCE$_{\scalebox{0.6}{\textsc{min}}}(c_3^{\scalebox{.6}{\textsc{{$\cal N$}}}}$,  $c_4^{\scalebox{.6}{\textsc{{$\cal N$}}}}$) = 767.503 nat, at the point $c_{3\,{\scalebox{0.6}{$\textsc{gmin}$}}}^{\scalebox{.6}{\textsc{{$\cal N$}}}}=1.264$ and $c_{4\,{\scalebox{0.6}{$\textsc{gmin}$}}}^{\scalebox{.6}{\textsc{{$\cal N$}}}}=0.169$, are in good agreement with the ones from Ref. \cite{Gutsche:2012wb} within 1.1\% and 5.6\%, respectively, and for the Roper resonance the value of
DCE$_{\scalebox{0.6}{\textsc{min}}}(c_3^{\scalebox{.6}{\textsc{$\cal R$}}}$,  $c_4^{\scalebox{.6}{\textsc{$\cal R$}}}$) = 17.639 nat, at the point $c_{3\,{\scalebox{0.6}{$\textsc{gmin}$}}}^{\scalebox{.6}{\textsc{{$\cal R$}}}}=0.803$ and $c_{4\,{\scalebox{0.6}{$\textsc{gmin}$}}}^{\scalebox{.6}{\textsc{{$\cal R$}}}}=-0.171$, are in good agreement with the corresponding data from Ref. \cite{Gutsche:2012wb} within 0.57\% and 2.43\%, respectively.
The obtained parameters imply that the calculations via DCE show the natural choice of the main properties of the resonance as the optimal tool for describing the localized nuclear system, which points to the most stable configuration of the system.
It should be stressed out also that in such a state the nuclear system can be represented as a set of data at a high compression rate of information through the wave modes describing the spatial complexity of the system.

Our calculation shows that behind the range in the plot in Figs. \ref{plot1} and \ref{plot2}, there are six another local minima (\textsc{lmin}) for nucleon resonance. The first one has DCE given by 
\begin{eqnarray}
\text{DCE$\left(c_{3\,{\scalebox{0.6}{$\textsc{lmin1}$}}}^{\scalebox{.6}{\textsc{{$\cal N$}}}}, c_{4\,{\scalebox{0.6}{$\textsc{lmin1}$}}}^{\scalebox{.6}{\textsc{{$\cal N$}}}}\right)$ = 662.012 nat},
\label{dce20}
\end{eqnarray}
at the point of the parameter space given by
\begin{eqnarray}
 \text{$c_{3\,{\scalebox{0.6}{$\textsc{lmin1}$}}}^{\scalebox{.6}{\textsc{{$\cal N$}}}}=0.125$,\qquad\qquad\qquad $c_{4\,{\scalebox{0.6}{$\textsc{lmin1}$}}}^{\scalebox{.6}{\textsc{{$\cal N$}}}}= 0.476$.}
\label{dce220}
\end{eqnarray}
The second local minimum is given by 
\begin{eqnarray}
\text{DCE$\left(c_{3\,{\scalebox{0.6}{$\textsc{lmin2}$}}}^{\scalebox{.6}{\textsc{{$\cal N$}}}}, c_{4\,{\scalebox{0.6}{$\textsc{lmin2}$}}}^{\scalebox{.6}{\textsc{{$\cal N$}}}}\right)$ = 878.379 nat},
\label{dce21}
\end{eqnarray}
at
\begin{eqnarray}
 \text{$c_{3\,{\scalebox{0.6}{$\textsc{lmin2}$}}}^{\scalebox{.6}{\textsc{{$\cal N$}}}}=2.896$,\qquad\qquad\qquad $c_{4\,{\scalebox{0.6}{$\textsc{lmin2}$}}}^{\scalebox{.6}{\textsc{{$\cal N$}}}}= -3.429$.}
\label{dce221}
\end{eqnarray}
The third local minimum,
\begin{eqnarray}
\text{DCE$\left(c_{3\,{\scalebox{0.6}{$\textsc{lmin3}$}}}^{\scalebox{.6}{\textsc{{$\cal N$}}}}, c_{4\,{\scalebox{0.6}{$\textsc{lmin3}$}}}^{\scalebox{.6}{\textsc{{$\cal N$}}}}\right)$ = 890.056 nat},
\label{dce22}
\end{eqnarray}
at
\begin{eqnarray}
 \text{$c_{3\,{\scalebox{0.6}{$\textsc{lmin3}$}}}^{\scalebox{.6}{\textsc{{$\cal N$}}}}=0.011$,\qquad\qquad\qquad $c_{4\,{\scalebox{0.6}{$\textsc{lmin3}$}}}^{\scalebox{.6}{\textsc{{$\cal N$}}}}= -3.608$}
\label{dce222}
\end{eqnarray}
represents a point wherein the nucleon has a high degree of configurational instability. There is a fourth local minimum, 
\begin{eqnarray}
 \text{$c_{3\,{\scalebox{0.6}{$\textsc{lmin4}$}}}^{\scalebox{.6}{\textsc{{$\cal N$}}}}=0.137$,\qquad\qquad\qquad $c_{4\,{\scalebox{0.6}{$\textsc{lmin4}$}}}^{\scalebox{.6}{\textsc{{$\cal N$}}}}= 2.916$;}
\label{dce223}
\end{eqnarray}
whose DCE reads 
\begin{eqnarray}
\text{DCE$\left(c_{3\,{\scalebox{0.6}{$\textsc{lmin4}$}}}^{\scalebox{.6}{\textsc{{$\cal N$}}}}, c_{4\,{\scalebox{0.6}{$\textsc{lmin4}$}}}^{\scalebox{.6}{\textsc{{$\cal N$}}}}\right)$ = 894.979 nat}.
\label{dce23}
\end{eqnarray}
Also, the fifth local minimum,
\begin{eqnarray}
 \text{$c_{3\,{\scalebox{0.6}{$\textsc{lmin5}$}}}^{\scalebox{.6}{\textsc{{$\cal N$}}}}=3.129$,\qquad\qquad\qquad $c_{4\,{\scalebox{0.6}{$\textsc{lmin5}$}}}^{\scalebox{.6}{\textsc{{$\cal N$}}}}= 2.138$;}
\label{dce224}
\end{eqnarray}
has associated DCE given by 
\begin{eqnarray}
\text{DCE$\left(c_{3\,{\scalebox{0.6}{$\textsc{lmin5}$}}}^{\scalebox{.6}{\textsc{{$\cal N$}}}}, c_{4\,{\scalebox{0.6}{$\textsc{lmin5}$}}}^{\scalebox{.6}{\textsc{{$\cal N$}}}}\right)$ = 918.743 nat},
\label{dce24}
\end{eqnarray}
Finally, the sixth local minimum of the DCE for the nucleon has DCE
\begin{eqnarray}
\text{DCE$\left(c_{3\,{\scalebox{0.6}{$\textsc{lmin6}$}}}^{\scalebox{.6}{\textsc{{$\cal N$}}}}, c_{4\,{\scalebox{0.6}{$\textsc{lmin6}$}}}^{\scalebox{.6}{\textsc{{$\cal N$}}}}\right)$ = 945.093 nat},
\label{dce25}
\end{eqnarray}
at the point of the parameter space,
\begin{eqnarray}
 \text{$c_{3\,{\scalebox{0.6}{$\textsc{lmin6}$}}}^{\scalebox{.6}{\textsc{{$\cal N$}}}}=0.705$,\qquad\qquad\qquad $c_{4\,{\scalebox{0.6}{$\textsc{lmin6}$}}}^{\scalebox{.6}{\textsc{{$\cal N$}}}}= 2.247$.}
\label{dce225}
\end{eqnarray}

For the Roper resonance, there are two local minima of the DCE. The first one,
\begin{eqnarray}
\text{DCE$\left(c_{3\,{\scalebox{0.6}{$\textsc{lmin1}$}}}^{\scalebox{.6}{\textsc{$\cal R$}}}, c_{4\,{\scalebox{0.6}{$\textsc{lmin1}$}}}^{\scalebox{.6}{\textsc{$\cal R$}}}\right)$ = 18.248 nat},
\label{dce26}
\end{eqnarray}
at
\begin{eqnarray}
 \text{$c_{3\,{\scalebox{0.6}{$\textsc{lmin1}$}}}^{\scalebox{.6}{\textsc{$\cal R$}}}=1.938$,\qquad\qquad\qquad $c_{4\,{\scalebox{0.6}{$\textsc{lmin1}$}}}^{\scalebox{.6}{\textsc{$\cal R$}}}= -1.762$;}
\label{dce226}
\end{eqnarray} is more stable than the second local minimum at
\begin{eqnarray}
 \text{$c_{3\,{\scalebox{0.6}{$\textsc{lmin2}$}}}^{\scalebox{.6}{\textsc{$\cal R$}}}=0.192$,\qquad\qquad\qquad $c_{4\,{\scalebox{0.6}{$\textsc{lmin2}$}}}^{\scalebox{.6}{\textsc{$\cal R$}}}= -2.987$.}
\label{dce227}
\end{eqnarray}
with DCE given by 
\begin{eqnarray}
\text{DCE$\left(c_{3\,{\scalebox{0.6}{$\textsc{lmin2}$}}}^{\scalebox{.6}{\textsc{$\cal R$}}}, c_{4\,{\scalebox{0.6}{$\textsc{lmin2}$}}}^{\scalebox{.6}{\textsc{$\cal R$}}}\right)$ = 21.750 nat},
\label{dce27}
\end{eqnarray}

One can see also the maximum in Fig. \ref{plot1} for the nucleon resonance with the value of
\begin{eqnarray}
\text{DCE$\left(c_{3\,{\scalebox{0.6}{$\textsc{gmax}$}}}^{\scalebox{.6}{\textsc{{$\cal N$}}}}, c_{4\,{\scalebox{0.6}{$\textsc{gmax}$}}}^{\scalebox{.6}{\textsc{{$\cal N$}}}}\right)$ = 1208.634 nat},
\label{dce3}
\end{eqnarray}
at the global maximum
\begin{eqnarray}
 \text{$c_{3\,{\scalebox{0.6}{$\textsc{gmax}$}}}^{\scalebox{.6}{\textsc{{$\cal N$}}}}=1.324$,\qquad\qquad $c_{4\,{\scalebox{0.6}{$\textsc{gmax}$}}}^{\scalebox{.6}{\textsc{{$\cal N$}}}} = -4.187$,}
\label{dce33}
\end{eqnarray}
 and five local maxima. The first local maximum, at 
 \begin{eqnarray}
 \text{$c_{3\,{\scalebox{0.6}{$\textsc{lmax1}$}}}^{\scalebox{.6}{\textsc{{$\cal N$}}}}=0.273$,\qquad\qquad $c_{4\,{\scalebox{0.6}{$\textsc{lmax1}$}}}^{\scalebox{.6}{\textsc{{$\cal N$}}}} = -1.934$;}
\label{dce44}
\end{eqnarray}
in the $(c_{3}^{\scalebox{.6}{\textsc{{$\cal N$}}}}, c_{4}^{\scalebox{.6}{\textsc{{$\cal N$}}}})$-parameter space has DCE given by 
 \begin{eqnarray}
\text{DCE$\left(c_{3\,{\scalebox{0.6}{$\textsc{lmax1}$}}}^{\scalebox{.6}{\textsc{{$\cal N$}}}}, c_{4\,{\scalebox{0.6}{$\textsc{lmax1}$}}}^{\scalebox{.6}{\textsc{{$\cal N$}}}}\right)$ = 1012.744 nat}.
\label{dce4}
\end{eqnarray}
It is slightly less stable than the second local maximum,  \begin{eqnarray}
 \text{$c_{3\,{\scalebox{0.6}{$\textsc{lmax2}$}}}^{\scalebox{.6}{\textsc{{$\cal N$}}}}=0.811$,\qquad\qquad $c_{4\,{\scalebox{0.6}{$\textsc{lmax2}$}}}^{\scalebox{.6}{\textsc{{$\cal N$}}}} = -0.847$,}
\label{dce55}
\end{eqnarray}
whose associated DCE is given by 
\begin{eqnarray}
\text{DCE$\left(c_{3\,{\scalebox{0.6}{$\textsc{lmax2}$}}}^{\scalebox{.6}{\textsc{{$\cal N$}}}}, c_{4\,{\scalebox{0.6}{$\textsc{lmax2}$}}}^{\scalebox{.6}{\textsc{{$\cal N$}}}}\right)$ = 1006.134 nat}.
\label{dce5}
\end{eqnarray}
Besides, the third local maximum \begin{eqnarray}
 \text{$c_{3\,{\scalebox{0.6}{$\textsc{lmax3}$}}}^{\scalebox{.6}{\textsc{{$\cal N$}}}}=0.825$,\qquad\qquad $c_{4\,{\scalebox{0.6}{$\textsc{lmax3}$}}}^{\scalebox{.6}{\textsc{{$\cal N$}}}} = -2.950$;}
\label{dce66}
\end{eqnarray}
is marginally less stable than the first and second local maxima (\ref{dce44}, \ref{dce55}), since its DCE, vaguely lower than the respective (\ref{dce4}, \ref{dce5}), reads
\begin{eqnarray}
\text{DCE$\left(c_{3\,{\scalebox{0.6}{$\textsc{lmax3}$}}}^{\scalebox{.6}{\textsc{{$\cal N$}}}}, c_{4\,{\scalebox{0.6}{$\textsc{lmax3}$}}}^{\scalebox{.6}{\textsc{{$\cal N$}}}}\right)$ = 1005.634 nat}.
\label{dce6}
\end{eqnarray}
Lastly, there is the   
\begin{eqnarray}
\text{DCE$\left(c_{3\,{\scalebox{0.6}{$\textsc{lmax4}$}}}^{\scalebox{.6}{\textsc{{$\cal N$}}}}, c_{4\,{\scalebox{0.6}{$\textsc{lmax4}$}}}^{\scalebox{.6}{\textsc{{$\cal N$}}}}\right)$ = 1008.927 nat},
\label{dce7}
\end{eqnarray}
at the fourth local maximum 
\begin{eqnarray}
 \text{$c_{3\,{\scalebox{0.6}{$\textsc{lmax4}$}}}^{\scalebox{.6}{\textsc{{$\cal N$}}}}=2.765$,\qquad\qquad $c_{4\,{\scalebox{0.6}{$\textsc{lmax4}$}}}^{\scalebox{.6}{\textsc{{$\cal N$}}}} = -0.852$}
\label{dce77}
\end{eqnarray}
and 
\begin{eqnarray}
\text{DCE$\left(c_{3\,{\scalebox{0.6}{$\textsc{lmax5}$}}}^{\scalebox{.6}{\textsc{{$\cal N$}}}}, c_{4\,{\scalebox{0.6}{$\textsc{lmax5}$}}}^{\scalebox{.6}{\textsc{{$\cal N$}}}}\right)$ = 1007.109 nat},
\label{dce8}
\end{eqnarray}
at the fifth local maximum,
\begin{eqnarray}
 \text{$c_{3\,{\scalebox{0.6}{$\textsc{lmax5}$}}}^{\scalebox{.6}{\textsc{{$\cal N$}}}}=2.721$,\qquad\qquad $c_{4\,{\scalebox{0.6}{$\textsc{lmax5}$}}}^{\scalebox{.6}{\textsc{{$\cal N$}}}} = -2.942$.}
\label{dce88}
\end{eqnarray}
There is one local maximum for the Roper resonance
\begin{eqnarray}
\text{DCE$\left(c_{3\,{\scalebox{0.6}{$\textsc{lmax1}$}}}^{\scalebox{.6}{\textsc{$\cal R$}}}, c_{4\,{\scalebox{0.6}{$\textsc{lmax1}$}}}^{\scalebox{.6}{\textsc{$\cal R$}}}\right)$ = 30.841 nat},
\label{dce9}
\end{eqnarray}
at
\begin{eqnarray}
 \text{$c_{3\,{\scalebox{0.6}{$\textsc{lmax1}$}}}^{\scalebox{.6}{\textsc{$\cal R$}}}=0.817$,\qquad\qquad $c_{4\,{\scalebox{0.6}{$\textsc{lmax1}$}}}^{\scalebox{.6}{\textsc{$\cal R$}}} = -2.250$.}
\label{dce99}
\end{eqnarray}
wherein the nuclear system is most unstable, from the configurational entropic point of view.

\noindent
As one can see from Fig. \ref{plot4} for the Roper resonance, the dark center of the blue domain of closed curves, in the first
quadrant in Fig. \ref{plot4} for the Roper resonance, indicates the point (\ref{dce110}), which matches the global minimum of the DCE (\ref{dce10}).

As one can note, the $(c_3^{\scalebox{.6}{\textsc{ $\cal N$, $\cal R$}}}, c_4^{\scalebox{.6}{\textsc{$\cal N$, $\cal R$}}})$-parameter space can be divided into configurational isentropic subsectors.
In Figs. \ref{plot2} and \ref{plot4} the contour plot consists of the closed curves joining points of the parameter space corresponding to the equal value of the DCE with the gradient of the DCE being orthogonal to the contour lines of the configurational isentropic curves.
Comparison of Figs.  (\ref{plot2}, \ref{plot4}) respectively for the nucleon and the Roper resonances shows that the parameters $c_3^{\scalebox{.6}{\textsc{{$\cal N$}}}}$, $c_4^{\scalebox{.6}{\textsc{{$\cal N$}}}}$ and $c_3^{\scalebox{.6}{\textsc{$\cal R$}}}$, $c_4^{\scalebox{.6}{\textsc{$\cal R$}}}$ affect the value of the DCE
in the way that it becomes steeper when the contour lines are close together.
In general, the inner isentropic subsector, represented by the colder colors, has lower values of the DCE, contrary, the outer isentropic subsector (hotter colors) refers to the higher values of the DCE.
And the dark blue subsector is surrounded by the global minimum DCE${}_{\scalebox{0.6}{$\textsc{gmin}$}}$($c_3^{\scalebox{.6}{\textsc{$\cal N$, $\cal R$}}},c_4^{\scalebox{.6}{\textsc{{$\cal N$}, $\cal R$}}})$.

Our data confirmed the suggestion that any limited number of degrees of freedom can define the most dominant state(s) of the nuclear configuration upon the thermodynamic equilibrium.
We intend to continue our study of DCE in the context of the seminal Refs.  \cite{Correa:2015lla,Correa:2015vka,DaRocha:2019fjr}.

\section{Conclusions}

The theory of a soft-wall AdS/QCD model has been used to determine the electro-excitation of $N^*(1440)$ Roper resonance. In particular, the excited Fock states are characterized by the leading three-quark $(3q)$ component. Such a state allows for determining the main properties of Roper resonance.
Within a soft-wall AdS/QCD, the gauge-invariant includes a coupling of two fermion AdS fields with the same twist-dimension for the nucleon and the ${\cal N}^*(1440)$ resonance.
To obtain the adjustable parameters, which determine the Lagrangian density of the interacted nucleons, as well as the gauge vector field in the ${\cal N}^*(1440)$ resonances, the concept of DCE has been used at small $Q^2$.
 Both for the nucleon and the Roper resonance, the DCE has a global minimum that determines the parameters respectively in Eqs. (\ref{par1}, \ref{par2}) with  good accuracy.
Such points in the parameter space present the higher compression rate of the information of the nuclear system, defining the condition of stability of the nuclear configuration, with a lower value of the uncertainty.
The global minima of the DCE, obtained for the corresponding value at the parameter space, confirm the data (\ref{par1}, \ref{par2}) for the transversal and longitudinal helicity amplitudes of the ${\cal N}^*(1440)$ resonance \cite{Gutsche:2012bp,Gutsche:2012wb}, which indicates that using the DCE approach is an appropriate tool for the study the nuclear reactions at high energy domain.
The calculation of the parameters within a soft-wall AdS/QCD using the DCE approach determines the natural choice for the excited nuclear configuration and can be successfully applied to other nuclear composite systems.

\paragraph*{Acknowledgments:} GK thanks to The São Paulo Research Foundation -- FAPESP (grant No. 2018/19943-6) and UFABC, for the hospitality. 
\medbreak

\paragraph*{Declaration of competing interest.} The authors declare that they have no known competing financial interests or personal relationships that could have appeared to influence the work reported in this paper.

\paragraph*{Data Availability Statements:} the datasets generated during and/or analyzed during the current study are available from the corresponding author upon reasonable request.

\end{document}